\documentstyle[prd,aps,preprint]{revtex}
\begin{document}
\title{Radiation-reaction-induced
evolution of circular orbits of particles around Kerr Black Holes}
\author{Daniel Kennefick}
\address{Theoretical Astrophysics,
California Institute of Technology, Pasadena, California 91125, USA}
\author{Amos Ori}
\address{Department of Physics, Technion - Israel Institute
of Techology, Haifa, 32000, Israel}
\maketitle
\begin{abstract}
It is demonstrated that, in the adiabatic approximation, non-Equatorial
circular orbits of particles in the Kerr metric (i.e. orbits of constant
Boyer-Lindquist radius) remain circular under the influence of
gravitational radiation reaction. A brief discussion is given of
conditions for breakdown of adiabaticity and of whether slightly
non-circular orbits are stable against the growth of eccentricity.
\end{abstract}
\section{Introduction}
It has been shown some time ago
that a particle in a circular orbit around a non-rotating black
hole remains in a circular orbit under the influence of the
gravitational radiation
reaction arising from its orbital motion  \cite{AKOP}.
Although it has been suggested \cite{Ori} that the same
holds true for ``circular'' orbits (meaning orbits of constant
Boyer-Lindquist radius \cite{circ}) around rotating black holes,
up to now it has not been shown to beyond first post-Newtonian order
(see Ref. \cite{Fintan} for the post-Newtonian result, and
see also Refs. \cite{Fin} and \cite{Mino} for more recent work)
because of the
difficulty of dealing with the little understood ``Carter''
constant of the motion, $Q$.
To date no practical method has been developed for describing
the rate of change of this constant due to gravitational radiation
reaction for a generic orbit (see, however, Ref. \cite{qdot}),
and without knowing this the evolution of the orbit cannot be
predicted.

In this paper, we study the relation between the rates of change of the
three constants of the orbital
motion (the energy, $E$, axial component of angular momentum, $L$,
and the Carter constant, $Q$), for circular and almost-circular orbits.
We first show that, for orbits which are precisely circular (in a sense
which is well-defined within the adiabatic approximation), the rate of
change of $Q$ has just the value required for the orbit to evolve into a
new circular orbit. Then, we extend the analysis to almost-circular
orbits (to first order in the orbital eccentricity).
This analysis leads to the
result that in order for an initially-circular orbit to develop
non-zero eccentricity, the back-reaction force (evaluated for the
precisely-circular orbit) must resonate with the radial oscillations.
Since in the case of a precisely-circular orbit the periodicity of the
back-reaction force is determined solely by the $\theta$-motion, we are led
to the following conclusion. The only case in which an initially-circular
orbit will develop an eccentricity is when there is a certain resonance
between the (small-oscillation) radial motion and the $\theta$-motion.
More specifically, this resonance condition is
$T_\theta =2 nT_r$ , where $n$ is an integer, and $T_r$ and $T_\theta$
are the (averaged) periods of the radial motion and the angular motion,
correspondingly.

Ryan \cite{Fin} has recently examined circular orbits in the Kerr metric
numerically and found that the above resonance condition is never
satisfied (for all black-hole and orbital parameters). We are thus
led to the conclusion that orbits which are initially precisely
circular will remain circular upon radiation-reaction evolution. We
point out, however, that this conclusion does not address the issue of
{\it stability} against the growth of a small initial eccentricity
(this issue is
further discussed in Sec. V below).

This paper is organized as follows. In Section II we
define instantaneous circularity of an orbit in terms of the
instantaneous location and 4-velocity of the orbiting particle. We
then derive a relation
between the rates of change of $E$, $L$ and $Q$ for an
instantaneously circular orbit that is perturbed by an arbitrary force
and we show that this relation is precisely the one required for the
circular orbit to evolve into a new circular orbit. One might naively
interpret this result by itself as implying that
an initially circular orbit will
necessarily remain circular. However, in Section III we show that in
order to predict the full evolution of initially circular orbits, it is
necessary to carry the analysis (of the relation between the rates of
change of the three constants of motion) to first order in the instantaneous
eccentricity. This analysis, to first order in the eccentricity, is
presented in Section IV, with the conclusion
that initially circular orbits do, indeed, remain circular.
Finally, in Section V we give some concluding
remarks.

\section{Instantaneously Circular Orbits}

In this section we shall define instantaneous circularity
of an orbit and shall show that for any such
orbits $dQ/d\tau$ has just the right value so as to
leave the evolving  orbit circular. We shall show that
this is true for any arbitrary  force which acts on
the orbiting particle (in fact, this result is
precise, and is not limited to the adiabatic
approximation).

We take here the point of view that the radiative evolution
may be viewed as the consequence of some ``back-reaction
force'', which can be treated as any other external force
\cite{force}.
We shall therefore begin by constructing a general formal expression
for the evolution rate of all constants of (geodesic)
motion, due to an arbitrary external force. Let  $C$  denote the
constant of motion in question. In Kerr,  $C$  may stand
for either the energy  $E$, the azimuthal angular
momentum $ L $, or the Carter constant  $Q$
(or any combination of these
constants,  like e.g. the constant $D$ defined below). We first
express all these  constants explicitly as
functions of coordinates and
covariant components of four-velocity \cite{one}, that is,
\begin{equation}
C\equiv
C(x^\beta ,u_\alpha )\;.\label{1}
\end{equation}
For  $E$  and $ L $, we simply take
\begin{equation}
E=-u_t\quad
,\quad  L =u_\varphi \;.\label{2}
\end{equation}
(Throughout, we use the
standard Boyer-Lindquist coordinates,  $r,t,\theta ,
\varphi $). The corresponding explicit
expression for  $Q$  is slightly more complicated. We
could use the familiar expression based on the
$\theta$-motion in Kerr:
\begin{equation}
Q=u_\theta ^{\;2}+\cos
^2\theta \,[a^2(1-u_t^{\;2})+\sin ^{- 2}\theta
\,u_\varphi ^{\;2}]\;.
\end{equation}
We find it more convenient,
however, to construct the expression for  $Q$  from
the $r$-motion. It is straightforward to show that
\begin{equation}
Q=\Delta ^{-1}\left[ {E(r^2+a^2)-a L } \right]^2-( L
-aE)^2-r^2- \Delta u_r^2\; \label{3}
\end{equation}
(this follows,
for instance, from Equations (33.32b) and (33.33c) in Ref.
\cite{MTW}). Here,  $\Delta \equiv r^2-2Mr+a^2$, where
$M$ is the black hole's mass and $aM$ is its angular momentum. For
later convenience, we also write this equation in the
form
\begin{equation}
Q=H(r,E, L )-\Delta u_r^2\;, \label{3a}
\end{equation}
where
\begin{equation}
H(r,E, L )\equiv \Delta ^{-1} \left[ {E(r^2+a^2)-a L
} \right]^2
-  ( L -aE)^2-r^2\;. \label{3b}
\end{equation}
In view of
Eq.\ (\ref{2}), one readily sees that Eq.\ (\ref{3}) is just of the
desired form,  $Q=Q(x^\beta ,u_\alpha )$  (with the
simplification that the right-hand side does not
depend on $\theta$ or $u_\theta$).

When an external force is applied to the particle, $C$
will evolve with time. To calculate its rate of
change, we differentiate Eq.\ (\ref{1}) with respect to the
proper time $\tau$:
\begin{equation}
{{dC} \over {d\tau }}=u^\beta
C_{,\beta }+\sum\limits_\alpha  {{{\partial C} \over
{\partial u_\alpha }}{{du_\alpha } \over  {d\tau }}}
\;. \label{4}
\end{equation}
Now,
\begin{equation}
{{du_\alpha } \over {d\tau }}=
{{Du_\alpha } \over {D\tau }} +{\textstyle{1 \over
2}}g_{\mu \nu ,\alpha }u^\mu u^\nu
\end{equation}
  and
\begin{equation}
{{Du_\alpha } \over {D\tau }}=F_\alpha \;,
\end{equation}
  where
$D/D\tau$ denotes covariant proper-time differentiation and
$F_\alpha$ is the force per unit rest mass. Equation (\ref{4})
thus reads
\begin{equation}
{{dC} \over {d\tau }}=\left[ {u^\beta
C_{,\beta }+{\textstyle{1 \over 2}}\sum\limits_\alpha
{{{\partial C}  \over {\partial u_\alpha }} g_{\mu \nu
,\alpha }u^\mu u^\nu }} \right]+\sum\limits_\alpha
{{{\partial C}  \over {\partial u_\alpha }}F_\alpha
}\;. \label{5}
\end{equation}
Notice that the term in brackets does
not depend on the external force. When no external
force is applied,  $C$  is conserved. Thus, the term
in brackets must vanish identically. Equation (\ref{5})
therefore reads:
\begin{equation}
{{dC} \over {d\tau
}}=\sum\limits_\alpha  {{{\partial C} \over  {\partial
u_\alpha }}F_\alpha } \;. \label{6}
\end{equation}
 This is the
desired general expression for the evolution rate of
all constants of motion. Note that this is the {\it
precise} expression for the {\it instantaneous} rate of
change of $C$. We have not used the adiabatic
approximation (or any  other approximation) so far.

We now define an orbit to be {\it instantaneously circular} if
its instantaneous values of $E$, $L$ and $Q$ [defined by
Eqs.\ (\ref{2}) and (\ref{3a})] are precisely equal to those of
some circular geodesic orbit. At a moment when the orbit (on
which the force $F_\alpha$ acts) is instantaneously circular,
Eq.\ (\ref{3}) plus circularity ($u_r=0$) implies
\begin{equation}
{{\partial Q} \over
{\partial u_r }}=-2\Delta u_r=0\;.
\end{equation}
Inserting this into Eq.\ (\ref{6}) gives
\begin{equation}
{{dQ} \over {d\tau }}= {{\partial Q} \over {\partial
u_t }}F_t +{{\partial Q} \over {\partial u_\varphi
}}F_\varphi  \; . \label{7}
\end{equation}
Also, in view of
Eq.\ (\ref{2}), Eq.\ (\ref{6}) yields
\begin{equation}
{{dE} \over {d\tau }}=-F_t\quad
,\quad {{d L } \over {d\tau }}=F_\varphi  \; .
\label{8}
\end{equation}
 Substituting Eqs.\ (\ref{2}) and (\ref{8}) in Eq.\ (\ref{7}),
we obtain
\begin{equation}
{{dQ} \over {d\tau }}=Q_{,E}{{dE} \over
{d\tau }}+Q_{, L }{{d L } \over {d\tau }} \; .
\label{9}
\end{equation}
Finally, using Eq.\ (\ref{3a}), we rewrite
Eq.\ (\ref{9}) as
\begin{equation}
{{dQ} \over {d\tau }}=H_{,E}{{dE} \over
{d\tau }}+H_{, L }{{d L } \over {d\tau }} \; .
\label{10}
\end{equation}
This is our final expression for the actual,
momentary, rate  of change of $Q$ due to the external
force. [To avoid confusion , we emphasize that the
partial derivatives in the right-hand  side of
Eq.\ (\ref{10}) are to be calculated according to the explicit
expression (\ref{3b})].

We come now to the second part of this calculation,
that is to calculate the rate of change of $Q$
(compared to that of $E$ and  $L$) required for taking
a circular orbit into a new circular orbit. {}From
Eq.\ (\ref{3a}) we obtain
\begin{equation}
\Delta u_r^2=H(r,E, L )-Q\equiv
W(r,E, L ,Q) \; . \label{21}
\end{equation}
When applied to general geodesic orbits (with constant but arbitrary
$E$, $L$ and $Q$) this equation can be regarded as describing radial motion
in an effective potential $W(r,E,L,Q)$. Obviously that geodesic motion
is circular if and only if the particle sits at a radius $r$ where $W=0$
(or $u_r=0$) and where $W_,r=0$ (so the particle is at the minimum of the
effective potential). Correspondingly, an orbit on which a force $F_\alpha$
acts is instantaneously circular if and only if it instantaneously satisfies
\begin{equation}
W=0\quad ,\quad
W_{,r}=0 \; . \label{22}
\end{equation}
We now let the force $F_\alpha$ continue to act, but only for an
infinitesimal
proper time, $\delta \tau$. Following this action, the orbit will be
characterized by new parameters, $r^\prime$, $E^\prime$, $L^\prime$ and
$Q^\prime$. We denote the changes from the original parameters by $\delta$,
that is, $\delta r=r-r^\prime$, $\delta E=E-E^\prime$, etc. These changes
are infinitesimal because $F_\alpha$ is allowed to act for only an
infinitesimal time $\delta \tau$, before the orbit is once more examined for
circularity. The corresponding change in $W$ is given by
\begin{equation}
\delta
W=W_{,r}\delta r+W_{,E}\delta E+W_{, L }\delta  L
+W_{,Q}\delta Q \; . \label{23}
\end{equation}
In order for Eq. (18) to hold after the time $\delta \tau$ as well as
before (i.e. in order for the orbit to remain circular), we must demand
\begin{equation}
\delta W=0 \; . \label{23a}
\end{equation}
We denote the value of
$Q$ which corresponds to a circular orbit (for given $E$
and $L$) by $Q_{\rm circ}(E,L)$. Equations (\ref{23}) and (\ref{23a}) thus
yield, as a necessary condition for the orbit to remain circular
\begin{equation}
W_{,r}\delta r+W_{,E}\delta E+W_{, L }\delta
L +W_{,Q}\delta Q_{\rm circ}=0 \; , \label{23b}
\end{equation}
from
which $\delta Q_{\rm circ}$ is to be determined.
Now
Eq.\ (\ref{21}) implies
\begin{equation}
W_{,Q}=-1\quad ,\quad
W_{,E}=H_{,E}\quad ,\quad W_{,  L }=H_{, L } \; ,
\label{24}
\end{equation}
which together with $W_{,r}=0$ [Eq.\ (\ref{22})] reduces Eq.\ (\ref{23b})
to the form
\cite{ff}
\begin{equation}
\delta Q_{\rm circ}=
H_{,E}\delta E+H_{, L }\delta  L  \; . \label{25}
\end{equation}
Finally, dividing by the infinitesimal proper time lapse $\delta \tau$
and taking the limit $\delta \tau \rightarrow 0$, we obtain
\begin{equation}
{{dQ_{\rm circ}} \over {d\tau
}}=H_{,E}{{dE} \over {d\tau }} +H_{, L }{{d L }
\over{d\tau }} \; . \label{26}
\end{equation}
Comparing Eqs. (16) and (24), we arrive at the desired result. {\it At any
moment when the orbit, on which the arbitrary force $F_\alpha$ acts, is
instantaneously circular}, $F_\alpha$ produces
{\it an instantaneous evolution
of the orbit's Cater constant given by}
\begin{equation}
{{dQ}
\over {d\tau }}={{dQ_{\rm circ}} \over {d\tau }} \;,
\label{27}
\end{equation}
{\it which maintains instantaneous circularity.}

For later convenience, let us define
\begin{equation}
D\equiv Q_{\rm circ}-Q=D(E,L,Q) \; .
\label{30}
\end{equation}
Circular orbits are thus characterized by
$D=0$ . Equation (\ref{27})  then implies that if at a
particular moment $D=0$ , then
\begin{equation}
dD/d\tau =0\; .
\label{31}
\end{equation}
It should be emphasized again that all the
calculations done so far are {\it precise}, and do
not depend on the adiabatic approximation. Note also that
these calculations refer to the {\it instantaneous} rate
of change of the constants of motion, at a moment when
the orbit is {\it instantaneously} circular. As we shall
see in the next section,
Eq.\ (\ref{31}) by itself does {\it not} imply that an initially-circular
orbit will necessarily remain circular. One must go to the next order
in the eccentricity in order to derive this result.

\section{The Need for Eccentricity Corrections}

Equation (\ref{31}) (which holds whenever $D=0$, i.e.
whenever the orbit is instantaneously circular) has a
trivial exact solution, $D(\tau)=0$ . Does this
necessarily mean that an instantaneously circular orbit will
remain circular forever? We shall immediately
see that, in principle, the answer is no (though, we
shall also see later that, {\it within the adiabatic
approximation}, in most cases a circular orbit will
remain circular). In fact, the trivial solution
$D(\tau)=0$ to Eq.\ (\ref{31}) is physically meaningless.

To illustrate this, consider the analogous (but much
simpler) problem of a free particle in one-dimensional
Newtonian mechanics. The particle's energy
(per unit rest mass) is
 $K=(1/ 2)\left( {dx/ d\tau } \right)^2$ .
 Assume now that a constant external force $F$ (per
unit rest mass) is applied on the particle.  The
evolution of $K$ is then given by [in analogy with
Eq.\ (\ref{6})]
\begin{equation}
{{dK} \over {d\tau }}=F{{dx} \over {d\tau }}\; ,
\label{36}
\end{equation}
or, in terms of $K$ itself, by
\begin{equation}
{{dK} \over {d\tau }}=\sqrt 2 F K^{1/ 2} \;.
\label{37}
\end{equation}
Now, assume that the particle is initially at rest,
i.e. $K=0$ . Then, from Eq.\ (\ref{37}), $dK/d\tau=0$ .
Equation (\ref{37}) then admits a trivial exact solution,
\begin{equation}
K(\tau )=0 \;. \label{37a}
\end{equation}
This trivial solution, which means that the particle
will remain at rest forever, is obviously wrong.
Instead, the particle will certainly accelerate, and
the true physical solution is \cite{two}
\begin{equation}
K=(F^2/ 2)\tau ^2 \;. \label{38}
\end{equation}

The situation here with respect to Eq.\ (\ref{31}) and its
unphysical solution $D(\tau)=0$ is just analogous. As
will be shown below (Section IV), in the extension of
Eq.\ (\ref{31}) to slightly-eccentric orbits, $dD/d\tau$ is
(to the leading order) proportional to $D^{1/2}$ .
Then, in addition to the trivial solution $D(\tau)=0$,
there exists a non-trivial solution in which for
some short period $D$ grows like $\tau^2$ [analogous
to Eq.\ (\ref{38})], and this is the physical solution. This
implies that, momentarily, the eccentricity (which is
proportional to $D^{1/2}$) will grow
linearly with $\tau$, even if initially it
vanishes precisely.

Recall, however, that we are not particularly
interested here in the momentary rate of change at a
specific point along the orbit. Rather, we are
interested in the effective {\it  long-term} evolution
of $D$ . Within the adiabatic approximation, this
long-term evolution is described by an equation of the
form
\begin{equation}
\dot D=S(E, L ,D) \;. \label{32}
\end{equation}
Hereafter, an overdot denotes the {\it  long-term}
rate of change (with respect to the proper-time, $\tau$), obtained
from the momentary equation of motion by averaging
over many periods (in Section IV we shall describe this
averaging procedure in more detail). The long-term
evolution of circular orbits will depend on the
asymptotic behavior of $S$ near $D=0$. It is not
difficult to show, based on Eq.\ (\ref{31}), that the zero-order
term [i.e. $S(D=0)$] vanishes identically \cite{zeroorder}.
As we shall see in Section IV
below, the general asymptotic form of $S$ is given by
\begin{equation}
S(E, L ,D)=S_1(E, L )D^{1/ 2}+S_2(E, L )D+O(D^{3/
2}) \;. \label{33}
\end{equation}
The evolution of an instantaneously circular orbit will depend
crucially on whether $S_1$ vanishes or not. In the
case $S_1=0$ , we can approximate Eq.\ (\ref{33}) by the
linear equation
\begin{equation}
\dot D=S_2D \;. \label{39}
\end{equation}
Then, an initial value $D=0$ ensures that $D$ will
remain zero forever (we are not concerned here about
stability to small initial perturbations, though we shall
discuss the question of stability briefly in Sec. V). On the
other hand, if $S_1$ is non-zero, we can approximate
Eq.\ (\ref{33}) by
\begin{equation}
\dot D=S_1D^{1/ 2} \;. \label{34}
\end{equation}
As was explained above, the trivial solution
$D(\tau)=0$ is physically meaningless. In that case, the physical
solution will be (at least as long as $S_1>0$)
\begin{equation}
D(\tau )=(S_1/ 2)^2\tau ^2 \;, \label{35}
\end{equation}
which means that the an instantaneously circular orbit will
evolve into an eccentric one.

The above considerations make it clear that the value
of $S_1$ is crucial for our discussion. An orbit which
is initially instantaneously circular will (in the adiabatic
approximation) remain circular if and only if $S_1=0$ .
In the next section we shall calculate $S_1$, and
show that it generically vanishes. Only orbits  which
satisfy a certain resonance condition may have
non-zero $S_1$ .

\section{Slightly Eccentric Orbits}

We now analyse the adiabatic evolution of $D$ for
slightly eccentric orbits, to leading order in the
eccentricity. In order to use Eq.\ (\ref{6}) for the
calculation of $dD/ d\tau $, we must first express
$D$ in the form $D(x^\beta ,u_\alpha )$. To simplify
the notation, we make use of Eq.\ (\ref{2}) , and simply
write $L$ instead of $u_\varphi$ and $-E$ instead of $u_t$ .
Recalling that  $Q_{\rm circ}\equiv Q_{\rm circ}(E,L)$, we
have from Eq.\ (\ref{30})
\begin{equation}
D=Q_{\rm circ}(E,L)-Q(r,u_r,E,L)=D(r,u_r,E,L) \;.
\label{110}
\end{equation}
Here, $Q(r,u_r,E,L)$ is to be understood in terms of
Eq.\ (\ref{3}). This function is analytic in its arguments,
and it is also not difficult to show that (except
perhaps at the ``last stable circular orbit'', which
must be excluded here), $Q_{\rm circ}(E,L)$ is analytic in
$E$ and $L$. Therefore, $D(r,u_r,E,L)$ in Eq.\ (\ref{110})
is also analytic in its arguments. Now, Eq.\ (\ref{6}) yields
\begin{equation}
{{dD} \over {d\tau }} =
-D_{,E}F_t+D_{,L}F_\varphi +D_{,u_r}F_r
 \equiv  D^t F_t+D^\varphi F_\varphi +D^r F_r = D^jF_j \;,
\label{120}
\end{equation}
where the index j runs over the three coordinates
$r,t,\varphi$ , and
\begin{equation}
D^j\equiv \partial D/ \partial u_j \;. \label{125}
\end{equation}
Equation (\ref{120}) [like Eq.\ (\ref{6})] describes the
{\it precise}, instantaneous rate of change of $D$, for any orbit
on which any force $F_\alpha$ is acting. We have
not made use of the adiabatic (or any other) approximation so far.

{}From the analyticity of $D(r,u_r,E,L)$ it follows that
the functions
\begin{equation}
D^j\equiv D^j(r,u_r,E,L)
\end{equation}
 are analytic
in their arguments. Now, from the validity of Eq.\ (\ref{31})
for an external force of any type, it follows that for
an instantaneously circular orbit
[i.e. when $u_r=0$ and $r=r_{\rm circ} (E,L)$], all three functions $D^j$
vanish. In other words, for any given $E$ and $L$,
\begin{equation}
D^j(r=r_0,u_r=0,E,L)=0 \;,
\end{equation}
where $r_0\equiv r_{\rm circ}(E,L)$ .
We now expand $D^j(r,u_r,E,L)$ around $(r=r_0,u_r=0)$.
In view of the analyticity of these functions, we have
\begin{equation}
D^j(r,u_r,E,L)=\delta
rD_{,r}^j+u_rD_{,u_r}^j+O(\delta
r^2,u_r^2,u_r \delta r) \;, \label{130}
\end{equation}
where $\delta r\equiv r-r_0$ , and the functions
$D_{,r}^j$ and $D_{,u_r}^j$ (which, again, are
analytic functions of $r,u_r,E,L$) are evaluated at
$(\delta r=0,u_r=0)$ .

We are now going to use two approximations (or
expansions): the adiabatic approximation, and the
small-eccentricity approximation. These two
approximations are unrelated, and should not be
confused with each other. The exact instantaneous equation
of motion of $D$ is [cf. Eq.\ (\ref{120})]
\begin{equation}
{{dD} \over {d\tau }}=D^jF_j \; . \label{140}
\end{equation}
The {\it adiabatic approximation means} that the external force
$F_j$ is assumed to be small, and the right-hand
side is to be evaluated to linear order in it.
The {\it small-eccentricity
approximation means} that the eccentricity is assumed
small and Eq.\ (\ref{140}) is evaluated to
first-order in it.

In view of the small-eccentricity approximation, the
higher-order terms in the right-hand side of Eq.\ (\ref{130})
can be omitted. Substitution in Eq.\ (\ref{140}) then yields
\begin{equation}
{{dD} \over {d\tau }}=\left[ {\delta r(\tau
)D_{,r}^j+u_r(\tau )D_{,u_r}^j} \right]F_j(\tau ) \; .
\label{150}
\end{equation}

Let us now examine the implications of the two
approximations used here on the expression in the
right-hand side. In view of the adiabatic
approximation, the term in brackets is to be evaluated
as if the constants of motion are fixed and the motion
is geodesic.  In the most general case,  $D_{,r}^j$
and $D_{,u_r}^j$  are (like $D$) functions of
$(r,u_r,E,L)$. Here, due to the adiabatic
approximation, we can fix $E$ and $L$ . Moreover,
since $\delta r$ and $u_r$  are already first-order in
the eccentricity, when evaluating  $D_{,r}^j$ and
$D_{,u_r}^j$  we may take $u_r=0$ and $r=r_0$ . Thus,
in Eq.\ (\ref{150}),  $D_{,r}^j$ and $D_{,u_r}^j$  are just
constants (which depend parametrically on $E$ and $L$) \cite{three}.

Turn now to evaluate $\delta r(\tau)$ and $u_r(\tau)$
in the right-hand side of Eq.\ (\ref{150}).  Like the entire
term in brackets, they are to be evaluated as if the
motion is geodesic (with fixed $E,L,D$). In view of
the small-eccentricity approximation, we need only
calculate $\delta r(\tau)$ and $u_r(\tau)$ to the
leading order in the eccentricity. We start from the
``effective-potential'' relation
\begin{equation}
\Delta u_r^2=W(r,E,L,Q) \; \label{160}
\end{equation}
[cf. Eq.\ (\ref{21})],
and, recalling that $g^{rr}=g_{rr}^{-1}=\Delta / \rho^2$ ,
we write it as
\begin{equation}
(dr/ d\tau )^2=(\Delta ^2/ \rho ^4)u_r^2=\rho ^{-
4}W\Delta\;. \label{170}
\end{equation}
Here,
\begin{equation}
\rho ^2\equiv r^2+a^2\cos^2\theta  \; . \label{180}
\end{equation}
In Eq.\ (\ref{170}), as it stands, the radial motion is coupled
to the $\theta$-motion, through $\rho$ . In order to
decouple the two motions, we define a new independent
variable $\lambda$ by
\begin{equation}
d\lambda / d\tau =\rho ^{-2} \; . \label{190}
\end{equation}
The radial equation of motion now becomes
\begin{equation}
(dr/ d\lambda )^2=W\Delta  \; , \label{200}
\end{equation}
in which the right-hand side is independent of
$\theta$ .

{}From Eqs.\ (\ref{21}) and (\ref{3b}), $W$ is an analytic function
of $(r,E,L,Q)$ . Writing $Q=Q_{\rm circ}(E,L)-D$ , and
recalling the analyticity of $Q_{\rm circ}(E,L)$ , $W$ can
be expressed as an analytic function of $(r,E,L,D)$ .
For an instantaneously circular orbit (i.e. for $r=r_0$ and
$D=0$), both $W$ and $W_{,r}$ vanish [cf Eq.\ (\ref{22})]. The
expansion of $W$ near instantaneous circularity must therefore be of
the form
\begin{equation}
W=\hat A D+\hat B\delta r^2+O(D^2,D\delta r,\delta
r^3) \; , \label{210}
\end{equation}
where $\hat A$ and $\hat B$ are some functions of $E$
and $L$ . Correspondingly, the expansion of the right-hand
side of Eq.\ (\ref{200}) will take the form
\begin{equation}
W\Delta =A D-B\delta r^2+O(D^2,D\delta r,\delta r^3)
\; , \label{220}
\end{equation}
where $A\equiv \hat A\Delta _0$ and $B\equiv -\hat
B\Delta _0$  , and where
\begin{equation}
\Delta _0\equiv \Delta (r=r_0(E,L)) \; . \label{230}
\end{equation}
Combining Eqs.\ (\ref{200}) and (\ref{220}), and restricting
attention to the leading-order eccentricity effect, we
obtain
\begin{equation}
(d\delta r/ d\lambda )^2=A D-B\delta r^2 \; .
\label{250}
\end{equation}
Note that $A$ and $B$ are some functions of $E$ and
$L$ only. Therefore, as explained above, they can be
regarded here as fixed parameters.

Equation (\ref{250}) describes a simple harmonic oscillator.
Its general solution is
\begin{equation}
\delta r(\lambda )=K\sqrt D \cos[\omega _r(\lambda -
\lambda
_0)] \; \label{280}
\end{equation}
where
$K=\sqrt {A/ B}$
and $\omega _r=\sqrt B$ are parameters that depend on
$E$ and $L$ only.
(Do not confuse $K$ here with $K$ of Sec. III).
Using this result to calculate $u_r$,
we find
\begin{equation}
u_r=g_{rr}{{dr} \over {d\tau }}=\Delta ^{-
1}{{d\delta r}
\over {d\lambda }}
=-\Delta ^{-1}K\sqrt D\omega _r
\sin[\omega
_r(\lambda -\lambda _0)] \; . \label{290}
\end{equation}
To simplify the notation, we shall hereafter absorb
the constant $\lambda _0$ into $\lambda$ (by shifting
the origin of the latter if necessary). Substituting
Eqs.\ (\ref{280}) and (\ref{290}) into Eq.\ (\ref{150}), we obtain
\begin{equation}
{{dD} \over {d\tau }}=\left[ K\sqrt D \right. \left(
 D_{,r}^j \cos(\omega _r\lambda ) \right.
- \left. \left. D_{,u_r}^j\Delta ^{-
1}\omega
_r \sin(\omega _r\lambda ) \right) \right]F_j(\tau )
\; . \label{295}
\end{equation}

{}From Eq.\ (\ref{295}) it is already clear that, so far as
the {\it instantaneous} evolution is concerned,
$dD/d\tau$ is indeed proportional to $\sqrt D$ . Thus,
as explained in Section III, although $D=0$ yields
$dD/d\tau=0$, an instantaneously circular orbit will not
remain circular later on. Instead, a
momentary growth of $D$ like $\tau^2$ is to be
anticipated. However, we are primarily interested here in
the {\it  long-term} adiabatic evolution of $D$ . The
latter is to be obtained from Eq.\ (\ref{295}) by averaging
it in $\tau$ over many periods. In order to perform
this long-term averaging, we must first take a closer
look at the nature of the time dependence of the force
$F_j$ .

Since the term in brackets on the right-hand side of
Eq.\ (\ref{295}) is already proportional to $\sqrt D$ (i.e.
to the eccentricity), when evaluating $F_j(\tau)$ we
are allowed to assume that the orbit is a precisely circular
geodesic. The $\theta$ motion of such an orbit is periodic in $\tau$.
Therefore, the backreaction force must be periodic
also. The various points along the circular geodesic orbit are
physically distinguishable from one another only by the values of $\theta$
and $d\theta/d\tau$. It therefore follows that
after completing a full cycle of the $\theta$-motion,
$F_j(\tau)$ will return to its original value. A
closer look, however, reveals that, because of the
reflection symmetry of the Kerr geometry, the
$\theta$-motion is symmetric with respect to the
equatorial plane. As a consequence, the period of
$F_j(\tau)$ will in fact be half of that of the full
$\theta$-motion cycle.

In order to facilitate the calculations, it is
convenient to transform Eq.\ (\ref{295}) from $\tau$ to
$\lambda$. Using Eq.\ (\ref{190}), we obtain
\begin{equation}
{{dD} \over {d\lambda }}=K\sqrt D \left[ \rho
^2 \left(
D_{,r}^j \cos(\omega _r\lambda ) \right. \right.
- \left. \left. D_{,u_r}^j\Delta ^{-
1}\omega
_r \sin(\omega _r\lambda ) \right) \right]
F_j(\tau(\lambda)) \; . \label{300}
\end{equation}
According to our expansion scheme, we need only
evaluate the term in brackets to zero order in the
eccentricity. That is, we can replace $\rho$ and
$\Delta$ by their circular counterparts, $\rho_0$
and $\Delta_0$ , where $\Delta_0$ is the constant
defined in Eq.\ (\ref{230}), and $\rho_0$ is a function of
$\theta$ only, defined by
\begin{equation}
\rho _0^2\equiv r_0^2+a^2 \cos^2\theta \;.
\label{305}
\end{equation}
 We therefore obtain
\begin{equation}
{{dD} \over {d\lambda }}=K\sqrt D \left( D_{,r}^j
\cos(\omega
_r\lambda ) \right.
- \left. D_{,u_r}^j\Delta _0^{-1}\omega _r
\sin(\omega
_r\lambda ) \right)\left[ {\rho _0^2 F_j}
\right] \; . \label{310}
\end{equation}
The term in brackets depends on $\tau$ (and hence on $\lambda$)
through its dependance on the $\theta$-motion.
It is obvious from Eq.\ (\ref{305}) that $\rho_0$ is periodic
in $\tau$, again with a period which is just half
that of the $\theta$-cycle. Consequently, the entire
term in brackets is also periodic (with that one-half
$\theta$ period). Let us examine now the periodicity of this
term with respect to $\lambda$. Again, the
 $\theta$-motion
is periodic in $\lambda$ , and the reflection
symmetry implies that the motion at $\theta<\pi/2$ is
just symmetric to that at $\theta>\pi/2$. (This can
also be deduced directly from the fact that, in the
equation of motion for $\theta(\lambda)$,  $(d\theta /
d\lambda )^2=\Theta (\theta )$ (cf. Ref. \cite{MTW}), the
function $\Theta (\theta )$ admits a reflection
symmetry about $\pi/2$). Thus, if we denote the
$\lambda$-period of the $\theta$-motion by
\begin{equation}
\Lambda _\theta \equiv 2\pi / \omega _\theta \; ,
\label{350}
\end{equation}
then the term in brackets in Eq.\ (\ref{310}) has a
$\lambda$-period of $\Lambda _\theta / 2$ .
Correspondingly, the Fourier transform of this term
will take the form
\begin{equation}
\rho _0^2 F_j=\sum\limits_{n=0}^\infty
{G_j^ne^{i2n\omega _\theta \lambda }}+C.C. \; ,
\label{360}
\end{equation}
where $C.C.$ means the complex conjugate of the preceding term.

Substituting this expansion in Eq.\ (\ref{310}) yields
\begin{equation}
{{dD} \over {d\lambda }}=\sqrt D\left[ K \right. \left(
D_{,r}^j \cos(\omega _r\lambda ) \right.
- \left. \left. D_{,u_r}^j\Delta
_0^{-1}\omega _r \sin(\omega _r\lambda ) \right)
\right]\sum\limits_{n=0}^\infty  {G_j^ne^{2ni\omega
_\theta\lambda }}
+ C.C. \; . \label{365}
\end{equation}
It is convenient to transform the last expression from
Sine and Cosine to exponential functions.  The term in
brackets can be expressed as
\begin{equation}
K_+^je^{i\omega _r\lambda }+K_-^je^{-i\omega
_r\lambda }
\end{equation}
where
\begin{equation}
K_\pm ^j\equiv (K/ 2)\left( {D_{,r}^j\pm
iD_{,u_r}^j\Delta_0^{-1}\omega _r} \right) \; .
\end{equation}
Eq.\ (\ref{365}) then takes the form
\begin{equation}
{{dD} \over {d\lambda }}=\sqrt D\left(
K_+^je^{i\omega
_r\lambda } \right. + \left. K_-^je^{-i\omega _r\lambda }
\right)\sum\limits_{n=0}^\infty  {G_j^ne^{2ni\omega
_\theta\lambda }}
+ C.C. \; .
\end{equation}
Defining now
\begin{equation}
G_\pm ^n\equiv K_\pm ^jG_j^n \;,
\end{equation}
we obtain
\begin{equation}
{{dD} \over {d\lambda }}=\sqrt
D\sum\limits_{n=0}^\infty
\left( G_+^ne^{i(2n\omega _\theta +\omega _r)\lambda
} \right. + \left. G_-^ne^{i(2n\omega _\theta -\omega _r)\lambda }
\right)
+ C.C. \;. \label{380}
\end{equation}
Recall that in this equation the coefficients $G_\pm
^n$ depend on $E$ and $L$ , but not on $\lambda$ .

Equation (\ref{380}) describes (within the adiabatic
approximation, and to leading order in the
eccentricity) the instantaneous rate of change of $D$ . In
order to obtain from it the long-term rate of change,
we simply need to take its average over a sufficiently
long period of $\lambda$ . To that end, for any
function $U(\lambda)$ , we formally define the long-term
averaged rate of change
\begin{equation}
\left\langle {{{dU} \over {d\lambda }}}
\right\rangle \equiv
\mathop {\lim}\limits_{\Delta \lambda \to \infty
}{{\Delta
U} \over {\Delta \lambda }} \;, \label{385}
\end{equation}
where $\Delta U $ and $\Delta \lambda $ denote the
difference in $U$ and $\lambda$ , correspondingly,
between the two extremes of the $\lambda$-interval
considered. Although the averaging is over times long
compared to $1/\omega_\theta$ and $1/\omega_r$,
it is still short compared to the radiation reaction
timescale, which is the time for substantial orbital
inspiral. (Recall that since we are using the
adiabatic approximation here, if $U$ also depends on
the ``constants of motion'', they must be taken as fixed
constants in this averaging process.) The averaging of
the right-hand side of Eq.\ (\ref{380}) is trivial, in that the
term $e^{i(2n\omega _\theta \pm \omega _r)\lambda }$
will average to zero, unless $2n \omega _\theta \pm
\omega _r =0$ , in which case it averages to one. Since
both $\omega _\theta$ and $\omega _r$ are taken
to be positive, we need only worry about those terms
with $2n \omega _\theta - \omega _r$ in the exponent.
We thus obtain
\begin{equation}
\left\langle {{{dD} \over {d\lambda }}}
\right\rangle =\sqrt
D\sum\limits_{n=0}^\infty  {G_n\delta _{(2n\omega
_\theta-\omega_r)}} \;, \label{390}
\end{equation}
where $\delta$ is a function whose value is unity when
$2n\omega_\theta-\omega_r=0$ and zero otherwise, and
$G_n\equiv 2\,\Re \left( {G_-^n} \right)$, with $\Re$ meaning the
``real part of''.

At this stage it is already clear that, unless a
certain resonance condition is satisfied ($\omega
_r=2n\omega _\theta $ for some $n$), the right-hand
side of Eq.\ (\ref{390}) will vanish. Before we further
discuss the meaning and implications of this resonance
condition, however, we shall more directly connect our
result (\ref{390}) to the notation used in Section III, and
in particular to the parameter $S_1$ . Equation (\ref{34})
is to be obtained from the momentary rate of change of
$D$ by averaging over proper time. For any function
$U(\tau)$, the long-term proper-time average (denoted
by an overdot) may be formally defined as
\begin{equation}
\dot U \equiv  \mathop {\lim}\limits_{\Delta \tau \to
\infty}{{\Delta U} \over {\Delta \tau }}
= \mathop
{\lim}\limits_{\Delta \lambda \to \infty }{{\Delta
\lambda }\over {\Delta \tau }}\mathop {\lim}\limits_{\Delta
\lambda\to \infty }{{\Delta U} \over {\Delta \lambda
}}=J^{-1}\left\langle {{{dU} \over {d\lambda }}}
\right\rangle \;. \label{393}
\end{equation}
where
\begin{equation}
J\equiv \left\langle {{{d\tau } \over {d\lambda }}}
\right\rangle  \; \label{396}
\end{equation}
is a constant that depends on the orbit. Applying this
procedure to $D$, we obtain \cite{five}
\begin{equation}
\dot D=J^{-1}\left\langle {{{dD} \over {d\lambda }}}
\right\rangle =\sqrt DJ^{-1}\sum\limits_{n=0}^\infty
{G_n\delta _{(2n\omega _\theta -\omega _r)}} \; .
\label{400}
\end{equation}

Comparing now Eq.\ (\ref{400}) to Eq.\ (\ref{34}), we find
\begin{equation}
S_1=J^{-1}\sum\limits_{n=0}^\infty  {G_n\delta
_{(2n\omega_\theta -\omega _r)}} \;. \label{420}
\end{equation}
The implications of this result for the long-term evolution of
$D$ are obvious. There are two different cases:

a)  The resonant case: there exists an integer
$n$ such that $\omega _r=2n\omega _\theta $ . In that
case, we have
\begin{equation}
S_1=J^{-1}G_n \;, \label{430}
\end{equation}
which is likely to be nonzero in the general case.
Then, as discussed in Section III [cf Eq.\ (\ref{35})] , $D$
will grow like $\tau^2$, which means that the
eccentricity will grow linearly with $\tau$.

b)  The non-resonant case: there exists {\it no}
integer $n$ for which $\omega _r=2n\omega
_\theta $ . In that case, we simply have
\begin{equation}
S_1=0 \;, \label{440}
\end{equation}
and the equation of evolution will read
\begin{equation}
\dot D=S_2D+O(D^{3/ 2}) \;. \label{450}
\end{equation}
[cf Eq.\ (\ref{33})]. In this case an orbit which is initially {\it
precisely} circular will remain circular (within the
adiabatic limit). (The sign of $S_2$ will determine
the stability against growth of small initial
eccentricity).

For resonant orbits, we have
\begin{equation}
\Lambda _\theta =2 n\Lambda _r \; \label{460}
\end{equation}
where $\Lambda _\theta $ and $\Lambda _r $ are the
$\lambda$-periods of the $\theta$- and $r$-motions,
correspondingly.
It would sometimes be convenient to translate this
expression to $t$-periods. One finds that, not
surprisingly,
the resonance condition is
\begin{equation}
T_\theta =2 nT_r \;, \label{470}
\end{equation}
where $T _\theta $ is the $t$-period of the
$\theta$-motion,
and $T _r $ is the {\it  averaged} $t$-period of
the radial motions. [The radial motion, expressed in
terms of $t$ (or $\tau$), is quasi-periodic rather than
periodic, because it is modulated by the $\theta$-motion.
The $\theta$-motion itself is periodic in either $t$ or
$\tau$ --- first, because we are considering a circular
orbit here, and second, because the resonance
condition (\ref{460}) ensures that each time $\theta$
returns to its original value, $r$ does also (but not
vice versa).]

\section{Conclusion}
We have shown that, within the adiabatic approximation, an orbit which
is initially precisely circular will remain circular, under the action of
the radiation-reaction force. The only exception is if the orbit
satisfies the resonance condition
$T_\theta =2 nT_r$ , for some integer $n$, where $T_\theta$ is the
$\theta$-motion period and $T_r$ is the (averaged) period of the
small-oscillation radial motion. However, circular orbits around Kerr
never satisfy this resonance condition \cite{Fin}. We therefore
conclude that, within the adiabatic approximation, an orbit which is
initially circular will remain circular.

There are two caveats which should be mentioned here. First, no attempt
has been made to address the issue of stability against the growth of
a small initial eccentricity. This stability would depend on the sign of
the coefficient $S_2$ in Eq.\ (\ref{33}) above, which was not calculated
here.

Second, our conclusion that circular orbits must remain circular is only
valid within the adiabatic limit, i.e. in the limit $\mu/M\to 0$, where
$\mu$ is the mass of the small object.
In reality, since the ratio $\mu/M$ is always
finite, an initially circular orbit will develop some eccentricity. For
concreteness, consider a particle with $\mu \ll M$ which at some initial
stage (which we denote stage 1) moves along a circular orbit with
Boyer-Lindquist radius $r_1$. Later on, the orbit shrinks due to
radiation reaction, until (at stage 2) it passes through a new radius,
$r_2<r_1$. Then, for every finite $\mu/M$, one should expect non-zero
eccentricity to be present at stage 2. The above analysis, however,
implies that the eccentricity at stage 2 will decrease with $\mu$ (for
fixed $r_1$, $r_2$ and $M$), and will vanish at the limit $\mu/M\to 0$
(presumably like $\mu/M$) \cite{nonadiabatic}.

The small eccentricity of non-adiabatic origin mentioned above could in
principle seed an exponential growth of eccentricity if $S_2>0$. In such
a situation, an initially circular orbit may evolve into a very
non-circular one (even for small $\mu/M$). The feasibility of this
scenario depends, of course, on the relevant values of  $\mu/M$ and
on whether $S_2>0$ and also on the available range of $r$-values (over which
$S_2$ is positive).

In reality, we know that in the Newtonian limit the orbits become more
and more circular as they shrink due to radiation reaction, so we expect
$S_2$ to be negative as long as $r \gg M$ (recall that the Newtonian
approximation should hold at $r \gg M$ even if the black hole is spinning).
Consequently, the range over which $S_2$ might be positive is bounded.
We can therefore expect that if $\mu/M$ is sufficiently small (and if
$r_2$ is not too close to the ``last stable circular orbit'' \cite{LSCO}),
the instability will not have enough time to build up, and an initially-
circular orbit will indeed remain circular throughout the
inspiral, to a good approximation.

\section{Acknowledgements} We gratefully thank Kip Thorne and Fintan Ryan
for many stimulating discussions and very helpful criticism. This research
was supported in part by NASA grant NAGW-4268 and by NSF grant AST-9417371.


\begin{thebibliography}{Longname}
\bibitem{Ori} This conjecture was made by one of us (A. Ori) several
years ago.
\bibitem{circ} Such orbits have often been referred to as ``quasi-circular''
in the literature (see, for instance, Ref. \cite{MTW}), because
such an orbit in Kerr does not describe a circle, or any closed curve and
is not even confined, in general, to a single plane. However, for our
purposes, the term circular aptly implies the analogy we wish to
draw with the Keplerian orbit case (or the Schwarzschild case). That is,
the orbit is described by a constant radial co-ordinate, and that
characteristic is maintained under the influence of an adiabatic back
reaction force.
\bibitem{force} In the case of gravitational radiation reaction
(which is our chief concern here), although it is possible to use
this concept of back-reaction force, this force is gauge-dependent.
We do not expect this ambiguity to affect our results, for the
following reason: despite the ambiguity in the momentary back-reaction
force, the long-term adiabatic evolution must be gauge-independent.
Therefore, one can simply pick any gauge in order to calculate the
radiative evolution. Once the gauge is chosen, the back-reaction force
is defined unambiguously, and all the analysis of this paper is applicable.
Keep in mind, however, that since we will employ an argument based
on the symmetries of Kerr, we must be careful to select a gauge which
does not artificially violate these symmetries. The harmonic gauge is
an example of a ``good'' gauge for our purposes.
\bibitem{one} Note that this function  $C(x^\beta ,
u_\alpha )$  is not unique. This is obvious, for
example, from the fact that   $- 1=g^{\alpha
\beta}(x)u_\alpha u_\beta $ .  The function
$C(x^\beta ,u_\alpha )$ can thus be chosen upon
convenience.
\bibitem{ff} Taking the differential of the second condition
in Eq.\ (\ref{22}), $W_{,r}=0$, does not yield any additional
constraint on $\delta Q_{\rm circ}$ (instead it results in an
expression for $\delta r_{\rm circ}$).
\bibitem{two} Mathematically, this ambiguity in the
solution of Eq.\ (\ref{37}) with the initial condition $K=0$
is possible due to the fact that the term $K^{1/2}$ in
the right-hand side is not differentiable with respect
to the unknown $K$. Note that this mathematical
ambiguity does not reflect any ambiguity in the
solution of the {\it  fundamental} equation of motion
(i.e. the Newton equation, $d^2x/ d\tau ^2=F$). The
only ambiguity occurs in the equation for $K$ derived from
$d^2 x/d\tau^2=F$. The origin of this ambiguity is simple. In
order to derive the equation for $K$, the fundamental
equation of motion is multiplied by $dx/d\tau$, and
this entity is identically zero in the trivial non-physical
solution (\ref{37a}). Thus, although the trivial
solution (\ref{37a}) satisfies the equation for $K$,
it does not satisfy the original equation of
motion.
\bibitem{three} Of the six coefficients  $D_{,r}^j$ and
$D_{,u_r}^j$, three vanish identically ($D_{,r}^r$,
$D_{,u_r}^t$, and $D_{,u_r}^\varphi$) but the
other three are generically nonzero.
\bibitem{five}It is sometimes useful to have the
averaged radiative evolution rates in terms of
external time $t$ rather than $\tau$. The procedure for
averaging over $t$ is analagous to the above
averaging over $\tau$. The result is also similar. In
the right-hand side of Eq.\ (\ref{400}), one only needs to
replace J by the corresponding parameter, $J_t\equiv
\left\langle {{dt}/{d\lambda }} \right\rangle $ .
\bibitem{nonadiabatic}
The back-reaction force (per unit mass $\mu$) is proportional to $\mu$
(and the evolution time from stage 1 to stage 2 is proportional to
$1/\mu$). Now, while the force terms responsible for the adiabatic
shrinking of the orbit from radius $r_1$ to $r_2$ are linear in the
back-reaction force, the non-adiabatic effects are (at least) quadratic
in that force. The non-adiabatic terms are therefore suppressed by the
small factor $\mu/M$ (compared to the adiabatic terms).
\bibitem{LSCO}
In Schwarzschild, it has been found \cite{AKOP} that $S_2$ diverges at
the last stable circular orbit, due to the flattening of the effective
potential well. Intuitively, one should expect the same behavior in Kerr
as well.
\end{thebibliography}
\end{document}